# Title:

# Super-resolution imaging of nanoscale inhomogeneities in hBN-covered and encapsulated few-layer graphene

Author(s) and Corresponding Author(s)*:

Lina Jäckering*,[1,2], Konstantin G. Wirth[1,2], Lukas Conrads[1,2], Jonas B. Profe[3,4], Alexander Rothstein[5,6], Hristiyana Kyoseva[5], Kenji Watanabe[7], Takashi Taniguchi[8], Dante M. Kennes[4,9], Christoph Stampfer[2,5,6], Lutz Waldecker[2,5], Thomas Taubner[1,2]

jaeckering@physik.rwth-aachen.de

## Affiliations

[1]1st Institute of Physics (IA), RWTH Aachen University, 52074 Aachen, Germany

[2]Juelich-Aachen Research Alliance (JARA-FIT), 52425 Juelich, Germany

[3]Institute for Theoretical Physics, Goethe University Frankfurt, 60438 Frankfurt a.M., Germany

[4]Institute for Theory of Statistical Physics, RWTH Aachen University and JARA Fundamentals of Future Information Technology, 52062 Aachen, Germany

[5]2nd Institute of Physics, RWTH Aachen University, 52074 Aachen, Germany

[6]Peter Grünberg Institut (PGI-9), Forschungszentrum Jülich, 52425 Jülich, Germany

[7]Research Center for Electronic and Optical Materials, National Institute for Materials Science, 1-1 Namiki, Tsukuba 305-0044, Japan

[8]International Center for Materials Nanoarchitectonics, National Institute for Materials Science, 1-1 Namiki, Tsukuba 305-0044, Japan

[9]Center for Free Electron Laser Science, Max Planck Institute for the Structure and Dynamics of Matter, 22761 Hamburg, Germany

**Abstract**

Encapsulating few-layer graphene (FLG) in hexagonal boron nitride (hBN) can cause nanoscale inhomogeneities in the FLG, including changes in stacking domains and topographic defects. Due to the diffraction limit, characterizing these inhomogeneities is challenging. Recently, the visualization of stacking domains in encapsulated four-layer graphene (4LG) has been demonstrated with phonon polariton (PhP)-assisted near-field imaging. However, the underlying coupling mechanism and ability to image subdiffractional-sized inhomogeneities remain unknown. Here, we retrieve direct replicas and magnified images of subdiffractional-sized inhomogeneities in hBN-covered trilayer graphene (TLG) and encapsulated 4LG, enabled by the hyperlensing effect. This hyperlensing effect is mediated by hBN's hyperbolic PhP that couple to the FLG's plasmon polaritons. Using near-field microscopy, we identify the coupling by determining the polariton dispersion in hBN-covered TLG to be stacking-dependent. Our work demonstrates super-resolution and magnified imaging of inhomogeneities, paving the way for the realization of homogeneous encapsulated FLG transport samples to study correlated physics.

**Introduction**

Few layer graphene (FLG) exists in different stacking orders which are characterized by their distinct crystallographic arrangement and can coexist in a single flake. Their crystal structure defines the stacking-specific band structure and therefore leads to stacking-dependent electronic and optical properties.[1]

FLG flakes are usually encapsulated in materials like hexagonal boron nitride (hBN)[2] or tungsten diselenide ($WSe_2$)[3] because hBN and $WSe_2$ provide an atomically flat environment to the graphene and thus encapsulated graphene flakes show an increased electron/hole mobility.[2,4] Improving the FLG device quality enabled to study half- and quarter-metals[5], ferroelectric and spontaneous quantum Hall states[6], and superconductivity[7] in rhombohedral trilayer graphene (TLG) and superconductivity in twisted bilayer graphene[8]. However, the fabrication of encapsulated FLG devices can induce inhomogeneities within the FLG flake such as changes in stacking domains or line defects.[9–11] Within the fabrication process, mechanical stress and strain during stacking and heating may lead to transformation of stacking domains.[9] Since high-quality transport samples of homogeneous, single-domain FLG are essential to study correlated physics, the visualization of stacking domains and line defects after the encapsulation is highly demanded.

Conventionally, IR- or Raman spectroscopy are used to characterize the stacking orders within an FLG flake.[12,13] However, both techniques are diffraction-limited and therefore cannot provide information about the FLG's local, sub-micrometer-sized electronic structure. The characterization of FLG stacking domains with IR- spectroscopy is strongly aggravated when the FLG is covered with another material like hBN because reflection at the surface reduces the FLG's signal.

Nanoscale characterization and visualization of stacking domains can be achieved with scattering-type scanning near-field optical microscopy (s-SNOM)[14,15] that overcomes the diffraction limit and has a lateral resolution down to 10 nm.[16] In s-SNOM, laser light illuminates a sharp tip that is brought into proximity to the sample's surface. The tip acts as an optical antenna. It strongly enhances the electric fields and collects the light scattered from the sample. The scattered light is then detected interferometrically yielding the amplitude $s_n$ and phase $\varphi_n$. In this way, the sample's local optical properties are obtained.

S-SNOM can be used in two different ways to characterize and visualize stacking domains and topographic defects of uncovered FLG. On the one hand, it can launch and image surface plasmon polaritons in graphene[17,18] that are reflected at inhomogeneities in the topography or in the optical properties e.g. at line defects and domain boundaries.[19,20] Thus, by imaging propagating polaritons s-SNOM can indirectly visualize domain boundaries.[19–21] On the other hand, different stacking orders show distinct s-SNOM signals, and thus, s-SNOM can also directly visualize and characterize stacking domains.[19,21–29]

Hexagonal boron nitride, which is usually used to encapsulate the graphene for transport measurements, is a dielectric 2D material and hosts phonon polaritons.[30,31] In a heterostructure of single-layer graphene (SLG) on top of hBN, the hBN phonon polaritons have been observed to couple to the plasmon polaritons in SLG.[32] Recently, Liu et al.[33] showed that phonon polaritons in hBN can be used to image the stacking domains of encapsulated four-layer graphene (4LG). However, the underlying coupling mechanism of the phonon polaritons in hBN to FLG and its stacking orders has not been studied. Further, the capability to image subdiffractional-sized inhomogeneities remains elusive.

Here, we demonstrate super-resolution and magnified imaging of subdiffractional-sized inhomogeneities in hBN-covered TLG and encapsulated 4LG with near-field microscopy that we attribute to the hyperlensing effect. The hyperlensing effect is mediated by hBN's hyperbolic phonon polaritons that couple to FLG's plasmon polaritons. First, we show that large stacking domains of TLG although being covered with hBN can directly be visualized with near-field microscopy due to their distinct optical conductivities. Secondly, we quantify the coupling of hBN's phonon polaritons to the TLG by determining the stacking-dependent dispersion of those coupled polaritons at an hBN edge. Thirdly, we demonstrate that those coupled polaritons mediate the hyperlensing effect and thus enable super-resolution and magnified imaging of buried structures. Finally, we provide an application-related example of how the hyperlensing effect helps to gain insight into device quality, e.g. the domain relaxation due to stacking and identification of buried topographic defects.

# Main Text

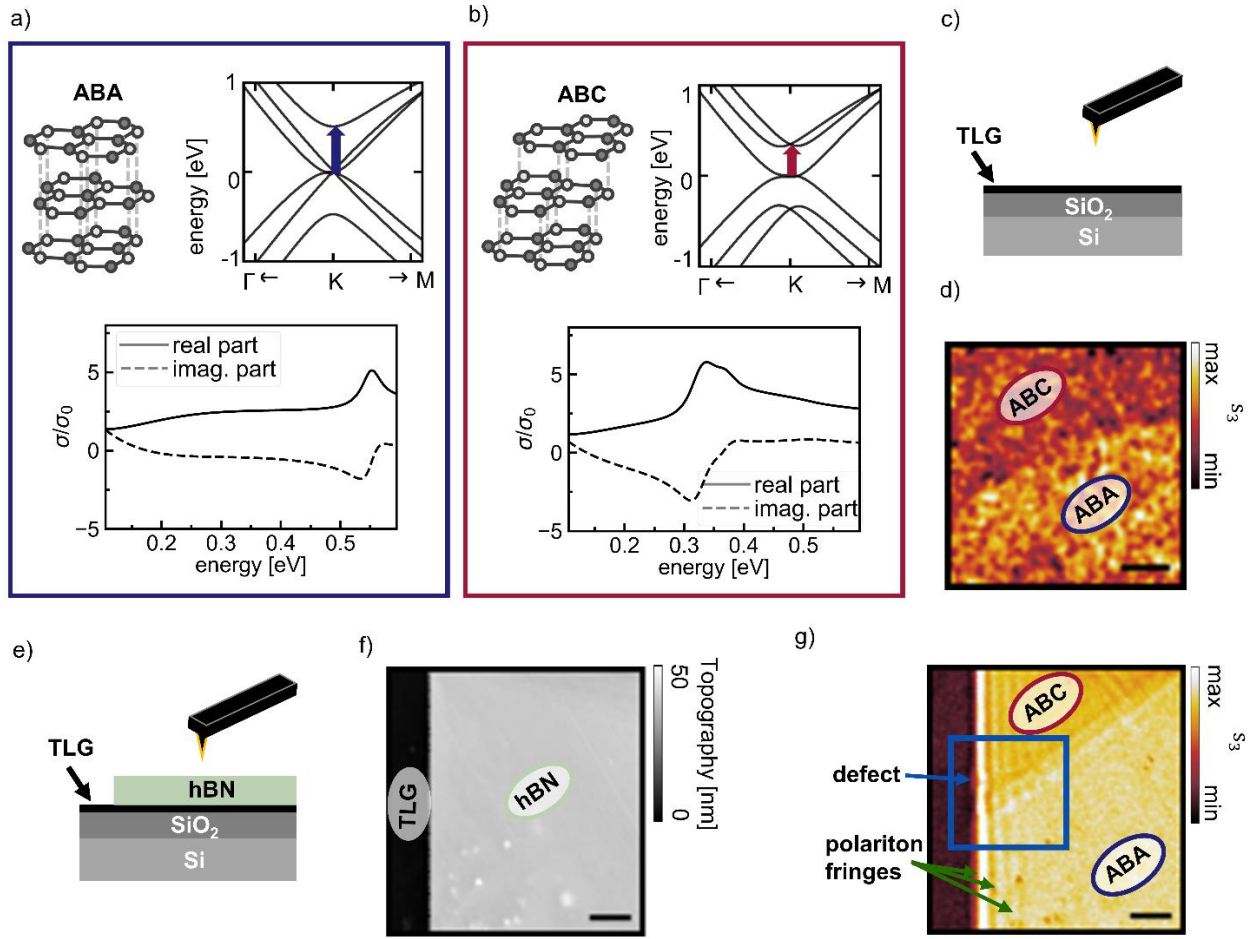

*Figure 1: s-SNOM imaging of stacking domains of uncovered and hBN-covered TLG. a) and b) Crystallographic stacking (top left), band structure (top right) and real and imaginary parts of the optical conductivity (bottom) of ABA and ABC stacked TLG, respectively. The blue and red arrows in the band structure indicate the interband transitions. c) Sketched side view of the investigated TLG on a $SiO_2$-Si substrate. d) The third order demodulated s-SNOM amplitude $s_3$ image taken at 0.176 eV shows two areas with different s-SNOM amplitudes that we assign to an ABC-stacked and an ABA-stacked domain, respectively. e) Sketched side view of the investigated heterostructure with 33 nm of hBN on top of TLG on a $SiO_2$-Si substrate. f) AFM-topography image of the heterostructure at an hBN edge. g) The s-SNOM amplitude $s_3$ image acquired a photon energy of at 0.181 eV along with the AFM-topography in f) shows an area of lower s-SNOM amplitude in the top corresponding to an hBN-covered ABC-stacked domain and an area of higher s-SNOM amplitude in the bottom corresponding to an hBN-covered ABA-stacked TLG. The green arrow marks the bright fringes parallel to the hBN edge that originate from the interference of polaritons. The blue rectangle marks the zoom-in discussed in Figure 3 to study the inhomogeneity marked by the blue arrow. The scale bars correspond to 1 µm.*

First, we use s-SNOM to characterize large stacking domains in uncovered and hBN-covered

TLG. TLG exists in two stacking orders, ABA and ABC, that differ in their arrangement of the graphene layers, see top left in Figure 1a and 1b, and have distinct electronic and optical properties. S-SNOM can be used to visualize areas with distinct optical properties, in general different permittivities. The optical properties of a 2D material are defined by the photon-energy-dependent optical conductivity which is connected to the optical permittivity: $\varepsilon = \varepsilon_0 + \frac{i\sigma\hbar}{E}$. The different crystallographic arrangement of the two TLG stacking orders results in different band structures (c.f. top right in Figure 1a and 1b). The different band structures result in different optical conductivities for the two stacking orders, as shown in the bottom in Figures 1a and 1b. The conductivities differ in the energy of the characteristic peak in the real part of the optical conductivity resulting from the resonant excitation of an interband transition, a transition between two electronic bands. While for ABA-stacked TLG the characteristic interband transition (blue arrow in the band structure) results in a peak of the real part of the optical conductivity at ~0.55 eV, for ABC-stacked TLG (red arrow in the band structure) it results in a peak of the real part of the optical conductivity at ~0.34 eV. Since the optical conductivities of the two TLG stacking orders differ in both their real and imaginary parts, the s-SNOM response is sensitive to the stacking order. We image an uncovered TLG flake as sketched in Figure 1c with s-SNOM. The s-SNOM amplitude image at 0.176 eV (c.f. Figure 1d) reveals two distinct areas with different amplitudes, which correspond to ABC- and ABA-stacked TLG domains. The assignment of the stacking domains is supported by Raman spectroscopy (see supplementary Figure S1). Thus, we show in agreement with previous s-SNOM imaging on uncovered FLG[19,21,22,24,27–29,34], that the stacking domains of TLG can be visualized by their distinct s-SNOM amplitudes which originate in their distinct optical conductivities.

For electronic transport measurements, FLG is usually encapsulated in hBN to increase the carrier mobility.[2] Therefore, we aim to characterize the stacking domains of hBN-covered TLG as sketched in Figure 1e. Imaging of buried structures up to a depth of 100 nm has been achieved with s-SNOM which uses exponentially decaying near-fields to investigate layered samples.[35–37] While the AFM topography image (Figure 1f) of an edge of the hBN flake on top of TLG shows a uniform thickness of 33 nm for the hBN flake, the simultaneously acquired s-SNOM amplitude image at 0.181 eV (Figure 1g) shows two distinct areas within the hBN flake with different s-SNOM amplitudes. We attribute the areas of different amplitudes to hBN-covered ABA and ABC TLG, respectively. The domain assignment is again supported by Raman spectroscopy (see supplementary Figure S1). Thereby, we show that the stacking domains can also directly be visualized by their distinct s-SNOM amplitudes when the TLG is covered with 33 nm of hBN.

Furthermore, we observe bright fringes parallel to the hBN edge (marked by the green arrows in Figure 1g) for hBN on both stacking TLG stacking orders. These fringes originate from the interference of polaritons launched by the tip and those reflected at the hBN edge. They are analyzed in detail in Figure 2 to elucidate the coupling of hBN's phonon polaritons to the TLG stacking orders. Moreover, we observe an inhomogeneity in the fringe pattern (blue arrow) in the hBN-covered ABC TLG domain. The origin of this inhomogeneity will be studied in detail in Figure 3 by analyzing zoomed-in s-SNOM images of the area marked by the blue rectangle.

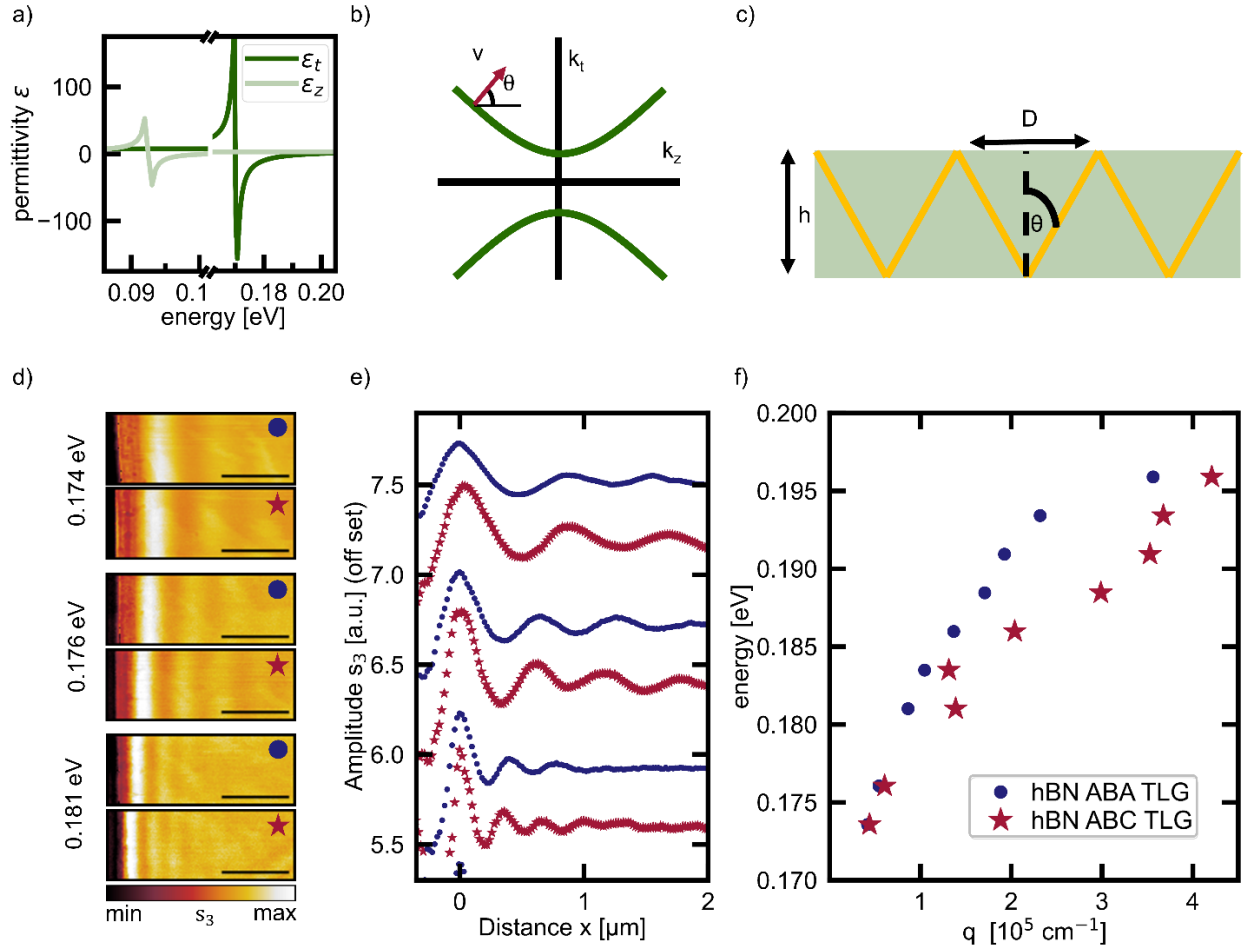

*Figure 2: Stacking dependent hyperbolic phonon plasmon polaritons: a) Real part of the in-plane (dark green) and out-of-plane (light green) permittivity of hBN taken from ref.[38]. b) Sketch of the hyperbolic isofrequency contours of hBN in its upper reststrahlenband. c) Sketched side view of the highly directional, propagating hyperbolic phonon polaritons in an hBN-slab. d) Third order demodulated s-SNOM amplitude images of the hBN-TLG heterostructure at an hBN edge for ABA (blue dot) and ABC (red star) stacked TLG for three different energies within the upper reststrahlenband of hBN. The scale bar is 1 µm. e) Line profiles extracted from the images in d) for both stacking orders, ABA as blue dots and ABC as red stars) and three different energies. f) Experimental dispersion relation of the hyperbolic phonon plasmon polaritons for the two stacking orders extracted from line profiles as exemplarily shown in e).*

Next, we elaborate on the underlying coupling of the phonon polaritons in hBN to the stacking orders of TLG. hBN is a highly anisotropic material because it has polar bonds in-plane and van-der-Waals bonds out-of-plane. Due to its anisotropy, it has two reststrahlenbands- the energy range between the longitudinal and the transverse optical phonon energy, where the real part of the permittivity is negative. Inside the investigated upper reststrahlenband (0.17-0.2 eV) the in-plane permittivity (dark green) is negative while the out-of-plane permittivity (light green) is positive as shown in Figure 2a. The opposite sign of the in-plane and out-of-plane permittivity leads to the hyperbolic isofrequency contours sketched in Figure 2b. Therefore, hBN is a natural hyperbolic material.[30,31,39] Due to its hyperbolic nature hBN hosts hyperbolic phonon polaritons which are volume-confined and highly directional.[39] At a fixed photon energy the hyperbolic phonon polaritons propagate under a constant angle and undergo multiple reflections at the interface of the hBN-slab as sketched in Figure 2c. In the limit of high wavevectors we can determine the propagation angle from the dispersion relation: $\tan\theta(E) = i\frac{\sqrt{\varepsilon_t(E)}}{\sqrt{\varepsilon_z(E)}}$.[39,40] The propagation angle is determined by the energy-dependent in-plane and out-of-plane permittivity and therefore can be tuned by varying the photon energy.

To characterize the polaritons in the hBN TLG heterostructures we recorded s-SNOM amplitude images at the hBN edge in Figure 1g for both stacking orders. Figure 2d shows the third order demodulated s-SNOM amplitude images for three different energies. At each energy two images are recorded, one for the hBN ABA TLG heterostructure marked by a blue dot and one for the hBN ABC TLG heterostructure marked by a red star. The s-SNOM amplitude images show bright fringes parallel to the hBN-edge that decrease in their amplitude away from the hBN edge. These bright amplitude fringes originate from the interference of tip-launched and edge-reflected propagating polaritons in the hBN slab.

From the line profiles in Figure 2e which are extracted from the s-SNOM images in Figure 2d, the wavelength of the polaritons can be determined. The spacing of the maxima corresponds to half of the polariton wavelength (arrow in Figure 2d) as the tip-launching mechanism dominates and edge-launching can be neglected here. We calculate the polariton wavevector $k_p = \frac{2\pi}{\lambda_p}$ from the extracted polariton wavelength and thereby determine the experimental dispersion for both heterostructures (hBN ABA TLG in blue and hBN ABC TLG in red in Figure 2f). The polariton wavevector (wavelength) increases (decreases) with increasing photon energy for both heterostructures. Comparing the propagation properties of the two heterostructures at the same energy, we observe three differences: First, the dispersion of the hBN ABC TLG heterostructure (red stars in Figure 2c) is shifted to higher wavevectors (lower wavelengths) compared to the dispersion of the hBN ABA TLG heterostructure (blue dots in Figure 2c). Second, the first maximum in the amplitude of the hBN ABC TLG heterostructure is closer to the hBN edge. Third, at 0.181 eV we observe five amplitude maxima for the hBN ABC TLG heterostructure while we observe only three amplitude maxima for the hBN ABA TLG.

The distinct propagation properties in the two heterostructures are quantified in the experimental dispersion (Figure 2f) and suggest that the hyperbolic phonon polaritons of hBN couple to the stacking-dependent plasmon polaritons in TLG as it was observed by Dai et al. [32] for SLG and hBN. We support the stacking-dependent dispersion of the hybridized modes in the heterostructures with modeled dispersion qualitatively agreeing with the experimental results (see supplementary Figure S2b-S2d). This stacking-dependent propagation behavior results from the distinct optical conductivities of the two stacking orders discussed in Figure 1. In the investigated energy regime (0.17-0.2 eV, c.f. zoom-in of the optical conductivities in Figure 1a and 1b in Figure S2a), the ABC TLG shows a higher magnitude of the imaginary part of the conductivity than ABA TLG. Since a higher magnitude of the imaginary part of the optical conductivity corresponds to a higher real part of the permittivity which leads to a larger polariton wavevector, the coupled polaritons in the hBN ABC TLG heterostructure show a higher wavevector than those in the hBN ABA TLG heterostructure. The real part of the optical conductivity instead corresponds to the imaginary part of the permittivity and thus influences the damping of the coupled polaritons. Due to ABA TLG's higher real part of the optical conductivity the coupled polaritons in the hBN ABA TLG heterostructure are stronger damped and show less fringes than those in the hBN ABC TLG heterostructure. Our observed polaritons are thus assigned to be stacking-dependent hyperbolic phonon plasmon polaritons (HP$^3$) and will reveal the underlying coupling mechanism of the PhP-assisted imaging.

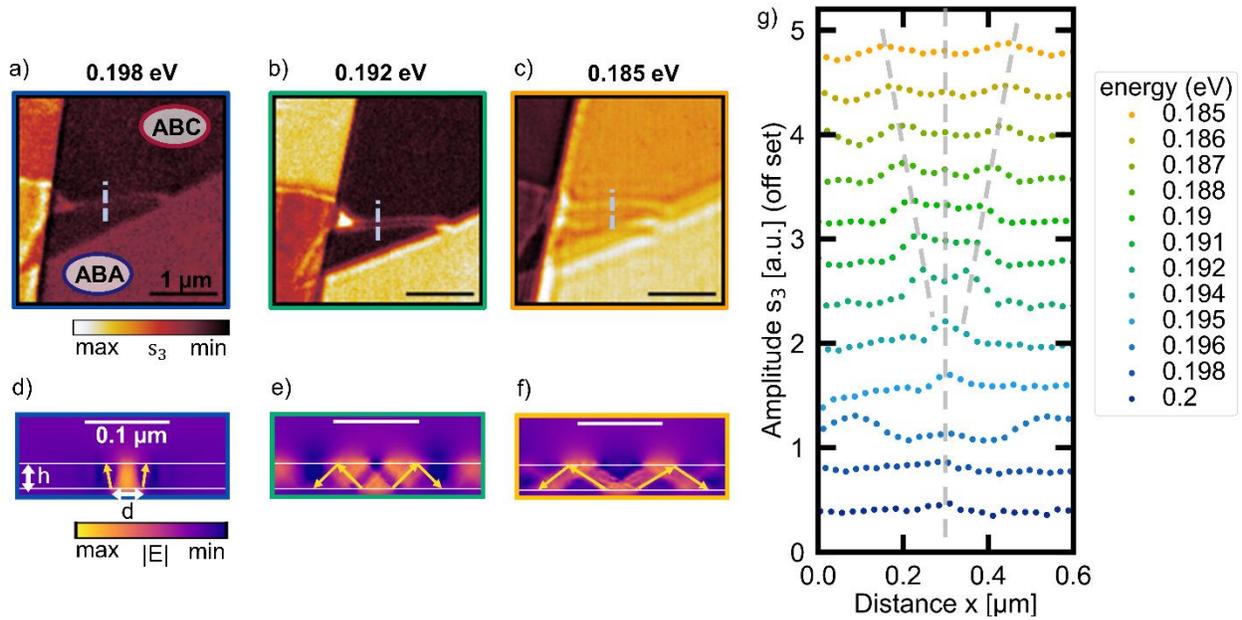

*Figure 3: Hyperlens-enabled inhomogeneity imaging in hBN-covered trilayer graphene. a)-c) Third order demodulated s-SNOM amplitude image of the hBN-TLG heterostructure with ABA-stacked TLG in the bottom and ABC-stacked TLG in the top at 0.198 eV, 0.192 eV, and 0.185 eV, respectively. Within the hBN-covered ABC-stacked TLG domain the amplitude images reveal a small structure of ABA-stacked domain. The scale bar is 1 µm. d)-f) Simulation of the electric field within the hBN-TLG heterostructure for the same energies as in a)-c). The scale bar corresponds to 0.1 µm. g) Line profiles along the blue dashed lines as exemplarily shown in a)-c) across the inhomogeneity within the ABC-stacked domain.*

In the following, we analyze the hyperbolic nature of the coupled polaritons in more detail and explain how they allow for subsurface imaging of a buried inhomogeneity. We study the area marked with the blue rectangle in Figure 1g where the polariton interference pattern at the hBN edge shows irregularities. Figures 3a-3c show the same sample area for three different energies within the upper reststrahlenband. At 0.198 eV there are areas of four different amplitudes. The two bright areas on the left correspond to uncovered TLG. The two areas on the right with a lower s-SNOM amplitude correspond to hBN-covered TLG. Within the hBN-covered TLG, we observe an area of lower s-SNOM amplitude at the top and an area of higher s-SNOM amplitude at the bottom. Based on the energy-dependent contrast, we assign the area with the higher s-SNOM amplitude to hBN-covered ABA TLG and the other area to hBN-covered ABC TLG. Within the hBN-covered ABC TLG domain there is a bright triangle of high amplitude close to the hBN-edge that continues as a bright line running from the hBN-edge to the boundary of the two

TLG stacking domains. We interpret this bright feature as a stacking inhomogeneity, specifically as a small domain of ABA-stacked TLG because it has a similar s-SNOM amplitude as the large domain of hBN-covered ABA TLG.

At 0.192 and 0.185 eV, we observe two bright fringes parallel to the line defect, which are broader and further apart for the lower energy of 0.185 eV. We interpret these fringes to originate from polaritons in the hBN slab launched at the buried inhomogeneity in the TLG. In agreement with the experimental dispersion (Figure 2c), the fringes are broader and further apart for lower energies because the polariton wavelength is longer for lower energies.

We can resolve a replica of the buried inhomogeneity at the hBN surface with an extension of below 300 nm although the used photon energies correspond to photon wavelength in the range of 6-7 µm. Thus, the 33 nm thick hBN layer allows to directly image subdiffractional-sized objects and therefore acts as a hyperlens.[41–43] The imaging of subdiffractional-sized objects through hBN is mediated by the phonon polaritons in hBN which have very high wavevectors due to the hyperbolic dispersion (see Figure 2b).

For a better understanding, we performed simulations (Figure 3d-3f) of the electric field across the cross-section of the hBN TLG heterostructure for the same three energies as in Figures 3a-3c. The hBN thickness h is assumed to be 33 nm and the small ABA TLG domain within the larger ABC TLG domain to have a width d of 40 nm. Upon illumination from the top, the ABA TLG domain launches polaritons due to the locally different optical conductivity. The analysis in Figure 2 showed that these polaritons are hyperbolic phonon plasmon polaritons. These HP$^3$ propagate under a restricted angle that is defined by the photon energy. At 0.198 eV the energy-dependent propagation angle is close to zero as sketched in Figure 3d by the yellow arrows. Thus, the HP$^3$ allow for a direct replica of the buried inhomogeneity in the TLG at the hBN surface. We achieve a super-resolution image similar to a near-field superlens.[44–46] For higher energies e.g. 0.192 eV and 0.185 eV, the propagation angle gets larger (yellow arrows in 3e and 3f) and we obtain a magnified image of the inhomogeneity in the TLG at the hBN surface.

We confirm the hyperlensing effect by analyzing line profiles across the inhomogeneity as indicated in Figure 3a-3c by dashed blue lines for several energies in Figure 3g. With decreasing energy (going from dark blue to orange) we observe that the single amplitude maximum in the line profile splits into two maxima whose distance increases, as marked by the grey dashed lines in Figure 3g. This behavior originates from the energy-dependent propagation angle of the HP[3]. With decreasing energy, the propagation angle increases leading to the formation of two maxima whose distance becomes larger. For high energies in the upper reststrahlenband, we observe a single maximum at the position of the inhomogeneity. This maximum gets broader with decreasing energy because the propagation angle increases; however, due to the finite resolution of the s-SNOM the two maxima cannot be resolved. At the photon energy of 0.192 eV the maximum splits into two maxima with a minimum in between at the position of the inhomogeneity. The angle is large enough such that the s-SNOM can resolve the two distinct maxima. For further decreasing energy the distance between the maxima increases. For energies smaller than 0.191 eV a third maximum is visible in the middle of the two maxima, at the inhomogeneity position. This behavior is in good agreement with previous results of hyperlensing with hBN conducted by Li et al.[39] and Dai et al.[40]. Both imaged subdiffractional-sized gold discs covered by hBN with s-SNOM and observed that the buried gold discs launch hyperbolic phonon polaritons. At a propagation angle of zero, a super-resolution image of the gold disc is obtained at the hBN surface whereas for higher propagation angles multiple rings are observed. Thereby, we confirm that the subsurface imaging is mediated by hyperbolic phonon plasmon polaritons. This is also in agreement with the recent study by Liu et al. [33] who attributed the imaging capabilities of domains in encapsulated 4LG to the hyperbolic nature of the phonon polaritons in hBN.

Moreover, in Figures 3b and 3c, we observe bright fringes parallel to the buried domain boundary originating from polaritons launched at the domain boundary. Thus, with the hyperlensing effect in hBN also buried domain boundaries can be visualized.

In addition, in Figure 3b the triangular-shaped inhomogeneity at the hBN edge shows a strongly enhanced s-SNOM amplitude. The enhancement is only observed for this specific energy and might originate from a resonator that is formed due to the geometry of the stacking domain. This strong enhancement already motivates to further study resonator structures for polaritons in hBN heterostructures. Especially tuneable resonators are of interest and could be realized by either gating when combing the hBN with graphene or by combining the hBN with phase change materials (PCMs). Combining hBN and the PCM IST, in which reconfigurable polariton resonators can be directly written with optical laser pulses[47,48], would allow to study arbitrarily shaped, reconfigurable polariton resonators without cumbersome fabrication techniques.

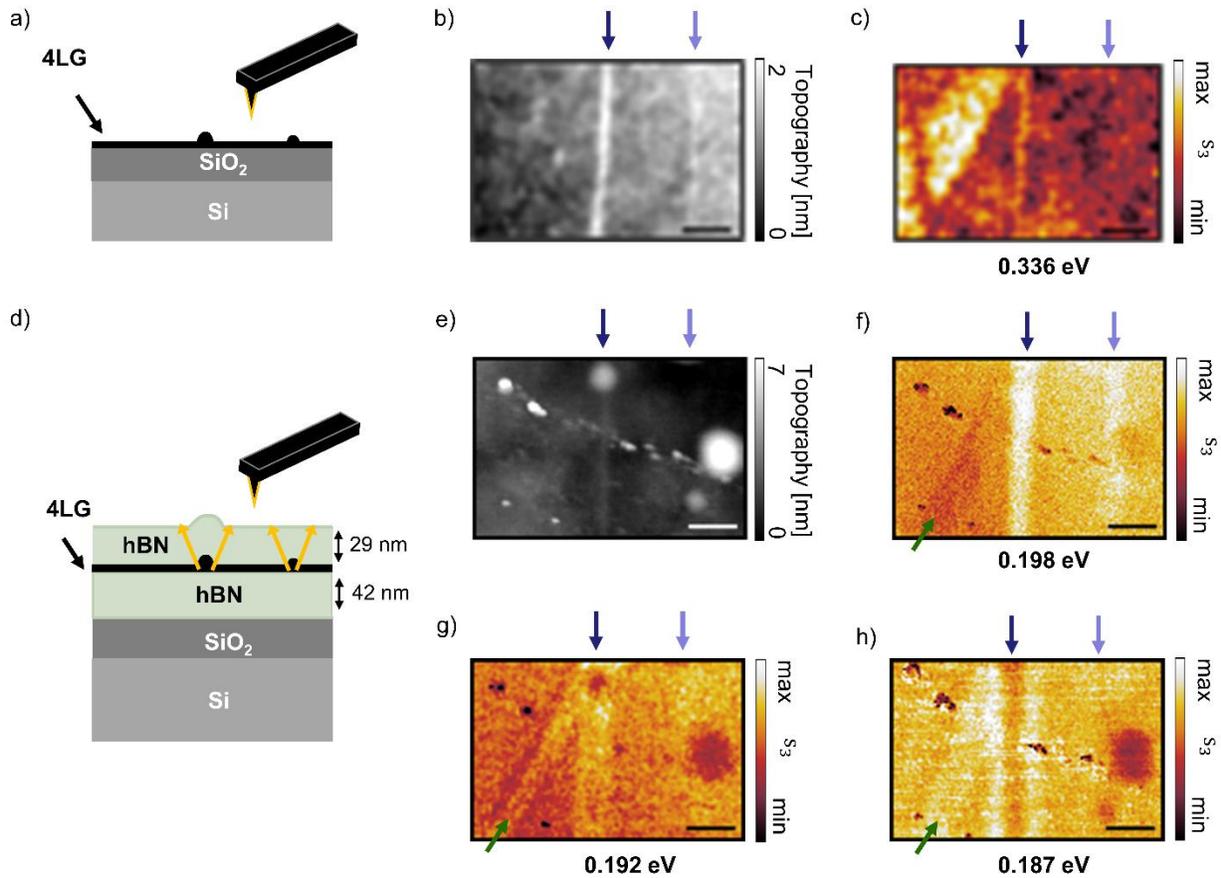

*Figure 4: Comparison of the domain structure of a 4LG flake before and after encapsulation with hBN. a) Sketched side view of the sample with 4LG on top of an SiO$_2$-Si substrate. b) AFM topography of the uncovered 4LG flake. c) Third order demodulated s-SNOM amplitude image of the 4LG flake at 0.336 eV revealing areas of two different amplitudes that correspond to the different 4LG stacking orders. d) Sketched side view of the same 4LG flake as in b) and c) after encapsulation and sketch of the polariton-mediated imaging of the inhomogeneities. e) AFM topography of the same 4LG flake as in b) and c) after encapsulation showing the same area as in b). f)-h) Third order demodulated s-SNOM amplitude image of the encapsulated 4LG flake at three different energies within the upper reststrahlenband of showing the same area as in c). The scale bars are 0.5 µm.*

We now apply super-resolution imaging mediated by the hyperbolic polaritons to visualize stacking domain relaxation and defect formation in a 4LG after encapsulation. Before encapsulation (see layer stack in Figure 4a), the 4LG flake was characterized by taking an AFM topography image (Figure 4b) simultaneously with an s-SNOM amplitude image (Figure 4c). We observe two bright vertical lines of higher topography (marked by the blue arrows) that may correspond to folds in the graphene flake. In the s-SNOM amplitude image taken at 0.366 eV, the 4LG flake shows two areas of distinct s-SNOM amplitudes. In a previous publication[28], we showed that with s-SNOM spectroscopy the characteristic interband transitions of the 4LG stacking orders can be retrieved allowing the identification of the 4LG stacking orders. In the amplitude image in Figure 4c, the areas with the higher s-SNOM amplitude correspond to the ABCA stacking and the areas with the lower amplitude to the ABAB stacking.

We encapsulated this 4LG flake in hBN (see layer stack in Figure 4d) and again acquired an AFM topography image (Figure 4e) along with s-SNOM images at different energies (Figure 4f-4h) at the same position as in Figure 4b and 4c. Besides some inhomogeneities with high topography, we observe a pronounced vertical line of high topography (dark blue arrow). We assign this topographic line to arise from the more pronounced line defect (dark blue arrow) in the uncovered 4LG flake (Figure 4b) that is wrapped by the hBN as sketched in Figure 4d.

This line defect is also visible as a line of high s-SNOM amplitude in the s-SNOM amplitude image of the encapsulated 4LG flake at 0.198 eV (Figure 4f). We observe a second narrower line of high s-SNOM amplitude (light blue arrow) next to it which we assign to the second line defect in the uncovered 4LG (light blue arrow in Figure 4b). We explain the observation of these buried topographic defects to the hyperlensing effect of the hBN cover layer elaborated in Figure 3. The topographic defects launch propagating coupled polaritons in the hBN slab as sketched by the orange arrows in Figure 4d. At 0.198 eV, the coupled HP[3] have a propagation angle close to zero, thus we obtain a nearly direct replica of topographic defects in the graphene flake at the hBN surface.

In the s-SNOM amplitude images of the encapsulated 4LG flake at 0.192 eV (Figure 4g) the line defect marked by the dark blue arrow appears broader and at 0.187 eV (Figure 4h) we observe two parallel bright lines with a minimum in between. For decreasing energies, the propagation angle of the polaritons increases. At 0.192 and 0.187 eV, the narrower line defect is not visible. The imaging of this narrow defect at lower energies might be hindered by the diffraction limit of the polaritons in the slab. The polariton wavelength increases with decreasing energy and therefore the diffraction limit ($\frac{\lambda_p}{2}$) increases. If the line defect is smaller than half of the polariton wavelength at 0.192 and 0.187 eV, respectively, it could not be resolved by the hyperlensing effect.

At 0.198 eV (Figure 4f) the s-SNOM amplitude image shows a feature of lower s-SNOM amplitude in the left, marked by the green arrow. At this position, we observe two parallel lines of high amplitudes at 0.192 and 0.187 eV (Figure 4g and 4h, respectively). Due to its lower s-SNOM amplitude at 0.198 eV, this inhomogeneity could be a small domain of another 4LG stacking order that is visualized via the hyperlensing effect.

In the three s-SNOM amplitude images in Figure 4f-4h we do not observe areas of different amplitudes as before encapsulation in Figure 4c. Therefore, we assume that the stacking domains have relaxed to the most stable ABAB stacking order. The relaxation of ABCA and ABCB-stacked domains to ABAB-stacked have been observed previously for uncovered 4LG-flakes under ambient conditions[28] and during the illumination with laser powers above 10 mW[49]. Further, the relaxation of rhombohedral to Bernal stacking domains in FLG has been observed upon dry transfer onto hBN and upon the application of metal contacts.[9]

**Discussion**

We demonstrate super-resolution and magnified imaging of subdiffractional-sized inhomogeneities such as stacking domains and topographic line defects in hBN-covered TLG and encapsulated 4LG with scattering-type scanning near-field optical microscopy (s-SNOM). We show that large stacking domains of TLG - although being covered with hBN - can directly be visualized with near-field microscopy due to their distinct optical conductivities. The visualization of subdiffractional-sized stacking domains instead relies on the hyperlensing effect, the super-resolution and magnified imaging. The hyperlensing effect is mediated by hBN's hyperbolic phonon polaritons that couple to FLG's plasmon polaritons. We confirm the coupling of hBN's phonon polaritons to the TLG by determining the dispersion of those coupled polaritons at an hBN edge to be stacking-dependent. Afterwards, we demonstrate that those coupled polaritons mediate the hyperlensing effect by visualizing a subdiffractional-sized stacking domain in hBN-covered TLG. The hyperlensing effect enables to obtain super-resolution images, direct replica at the surface of the hBN, and magnified images of buried structures. By comparing s-SNOM amplitude images and topography of a 4LG flake before and after encapsulation in hBN, we demonstrate that the hyperlensing effect can visualize the relaxation of stacking domains as well as subdiffractional-sized inhomogeneities and thereby pave the way for high-resolution characterization FLG transport samples after encapsulation.

The stacking-dependent HP$^3$ itself open a playground for gate-tuneable polaritonics. It has been shown that the HP$^3$ in a heterostructure of hBN and SLG combine both the low losses due to the hBN and the gate tuneability of SLG.[32] The HP$^3$ wavelength has been found to be tuneable by applying a gate voltage and thereby modifying the fermi energy. This concept can be easily applied to our hBN TLG heterostructure, where the two stacking orders have different band structures and thus might show different responses to the gate voltage.

The hyperlensing effect of hBN could also be a potential imaging mechanism for local inhomogeneities of the twist-angle in twisted BLG (tBLG). Using magnified imaging the direct visualization of domain walls in the moiré-pattern in encapsulated low-angle tBLG has been realized recently.[50] Since the hyperlensing effect is sensitive to local variations in the optical conductivity we expect that also local inhomogeneities in the twist angle of encapsulated tBLG with larger twist angle can be visualized. For larger twist-angles the moiré pattern becomes too small such that the domain walls cannot longer be directly visualized. Local variations in the twist-angle are expected to result in local variations in the optical conductivity and should therefore be detectable with the hyperlensing effect.

Phonon polariton-assisted imaging is not restricted to heterostructures with hBN but is expected to also be observed in other hyperbolic 2D materials like α-$MoO_3$.[51–53] In any heterostructure with a hyperbolic material as the top layer, the phonon polariton-assisted imaging can be used to visualize topographic defects or changes in the optical properties. The phonon polariton-assisted imaging opens opportunities for visualization and monitoring of stacking domain relaxations in van-der-Waals heterostructures during the process of device fabrication. It offers a reliable tool for device characterization after fabrication. The phonon polariton-assisted imaging allowed Liu et al.[33] to characterize the stacking domains of 4LG after encapsulation and thus enabled to fabricate devices only containing the rhombohedral stacking order. Using those devices, they studied symmetry breaking due to Coloumb interaction present in the rhombohedral stacking order. The phonon-polariton assisted imaging allows to visualize and characterize nanoscale stacking domains and topographic defects after encapsulation of the van-der-Waals material. Thus, the actual stacking order of an encapsulated transport device can be monitored, and it can be ensured that a homogenous flake is used to study correlated physics.

## Methods

**Exfoliation.** We exfoliate few-layer graphene (graphite flake source: Naturgraphit GmbH) and hBN on standard Si/SiO2 wafers (oxide thickness: 90 nm). Suitable flakes for further processing are identified by an automated microscopy setup.[54] The stacking process follows a standard dry transfer technique using a polycarbonate/polydimethylsiloxane (PC/PDMS) droplet[55]. For the fabrication of the hBN/tri-layer graphene structure we pick-up the hBN flake at a temperature of approximately 90°C and place it on top of the target tri-layer graphene flake, where we subsequently bake it at 165°C. For the fabrication of the hBN/four-layer graphene/hBN sandwich we start with picking up an hBN flake at 90°C followed by the four-layer graphene pick-up at 40°C. This structure is then placed on a target hBN flake and baked at 165°C.

**s-SNOM.** We performed near-field imaging of polaritons in hBN-covered TLG and encapsulated 4LG using a commercially available s-SNOM (NeaSNOM neaspec GmbH) at photon energies between 0.17 and 0.21 eV with a liquid nitrogen-cooled Mercury Cadmium Tellurium detector. The used laser sources are commercially available quantum cascade lasers (MIRcat-QT Mid-IR Laser and QCL by DRS Daylight Solutions). Near-field images of the domains in the 4LG flake were acquired at 0.336 eV using a commercially available tunable OPO/OPA laser system (Alpha Module, Stuttgart Instruments) combined with a liquid nitrogen-cooled indium antimonide detector. We operate the s-SNOM in the pseudoheterodyne detection mode, to obtain amplitude and phase. The s-SNOM is operated at tapping amplitudes of 75-85 nm. We evaluated the third-order demodulated amplitude images.

**Numerical Field Simulations.** The field simulations were performed with the numerical simulation program CST Studio Suite. Here, the anisotropic permittivity of 33 nm hBN is modelled according to Figure 2a, and the conductivities of the ABA and ABC TLG are taken into account. For excitation, Floquet Mode Ports are chosen with periodic boundaries in lateral dimensions. The tip of the SNOM is not taken into account.

**Calculation of the optical conductivity.** We employ the Kubo formula to calculate the optical conductivity of ABA and ABC TLG shown in Figures 1a and 1b in the main text. The starting point for the calculations is the tight-binding Hamiltonian. The model considers the nearest-neighbor in-plane hopping with a hopping energy $\gamma_0$ = 3.16 eV and the nearest-neighbor out-of-plane hopping with a hopping energy $\gamma_1$ = 0.39 eV. A detailed description of the applied formalism is provided in ref[10]. We assumed room temperature (0.025 eV). We chose a broadening of 20 meV and a Fermi energy of -80 meV, both were chosen such that the best agreement with the experimental results is obtained.

**Code Availability**

The computer codes developed for this study are available from the corresponding author upon request.

29

**Acknowledgements**

L.J. acknowledges the support of RWTH University through the RWTH Graduate Support scholarship. A.R., C.S., and L.W. acknowledge support from the European Union's Horizon 2020 research and innovation programme under grant agreement no. 881603 (Graphene Flagship), the Deutsche Forschungsgemeinschaft (DFG, German Research Founda-tion) under Germany's Excellence Strategy - Cluster of Excellence Matter and Light for Quantum Computing (ML4Q) EXC 2004/1-390534769, the FLAG-ERA grant PhotoTBG, by the Deutsche Forschungsgemeinschaft (DFG, German Research Foundation) - 471733165. K.W. and Ta.Ta. acknowledge support from the JSPS KAKENHI (Grant Numbers 20H00354 and 23H02052) and World Premier International Research Center Initiative (WPI), MEXT, Japan.

**Author Contributions Statement**

L.J., K.G.W, and T.T. conceived the project with input by L.W. and C.S. A.R. and H.K. fabricated the samples. L.J. performed the s-SNOM measurements and theoretical dispersion calculation. K.G.W. supervised parts of the s-SNOM measurements. L.W. carried out and analysed the Raman measurements. L.J. and K.G.W. analysed the experimental data. L.C. performed the numerical field simulations. J.B.P. and D.M.K. carried out the theoretical calculation. K.W. and Ta.Ta. provided high-quality hBN crystals. All authors contributed to writing the manuscript.

**Competing Interests Statement**

The authors declare no competing financial interest.

# Supplementary Information:

# Super-resolution imaging of nanoscale inhomogeneities in hBN-covered and encapsulated few-layer graphene

Author(s) and Corresponding Author(s)*:

Lina Jäckering*,1,2, Konstantin G. Wirth[1,2], Lukas Conrads[1,2], Jonas B. Profe[3,4], Alexander Rothstein[5,6], Hristiyana Kyoseva[5], Kenji Watanabe[7], Takashi Taniguchi[8], Dante M. Kennes[4,9], Christoph Stampfer[2,5,6], Lutz Waldecker[2,5], Thomas Taubner[1,2]

jaeckering@physik.rwth-aachen.de

**Affiliations**

[1]1st Institute of Physics (IA), RWTH Aachen University, 52074 Aachen, Germany

[2]Juelich-Aachen Research Alliance (JARA-FIT), 52425 Juelich, Germany

[3]Institute for Theoretical Physics, Goethe University Frankfurt, 60438 Frankfurt a.M., Germany

[4]Institute for Theory of Statistical Physics, RWTH Aachen University and JARA Fundamentals of Future Information Technology, 52062 Aachen, Germany

[5]2nd Institute of Physics, RWTH Aachen University, 52074 Aachen, Germany

[6]Peter Grünberg Institut (PGI-9), Forschungszentrum Jülich, 52425 Jülich, Germany

[7]Research Center for Electronic and Optical Materials, National Institute for Materials Science, 1-1 Namiki, Tsukuba 305-0044, Japan

[8]International Center for Materials Nanoarchitectonics, National Institute for Materials Science, 1-1 Namiki, Tsukuba 305-0044, Japan

[9]Center for Free Electron Laser Science, Max Planck Institute for the Structure and Dynamics of Matter, 22761 Hamburg, Germany

**This PDF file includes:**

    **Supplementary Note 1: Assignment of the stacking orders in the TLG flake**

    **Supplementary Note 2: Stacking-dependent polaritons in hBN-covered TLG**

**Supplementary Note 1: Assignment of the stacking orders in the TLG flake**

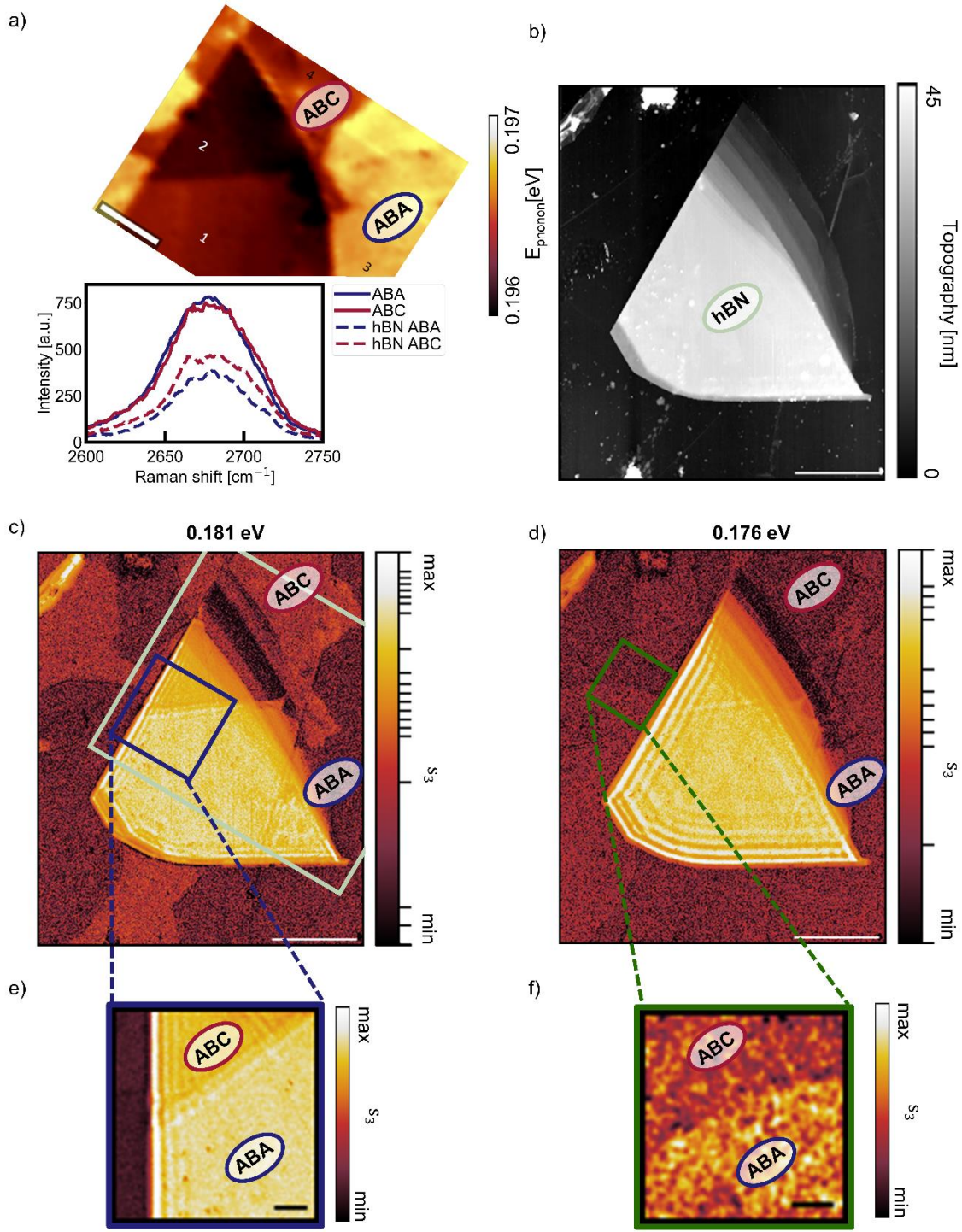

***Figure S1: Assignment of the stacking orders in the TLG flake.*** *a) Raman G-peak map (top) of the TLG flake partly covered with an hBN flake and Raman 2D-peak spectra for the four areas with distinct G-peak positions (bottom). The scale bar corresponds to 4 µm. b) AFM topography image of the same flake as in a) acquired simultaneously with the s-SNOM amplitude image in c). c) and d) s-SNOM amplitude images of the same hBN TLG heterostructure as in a) and b) at 0.181*

*and 0.176 eV, respectively. The light green rectangle in c) indicates the area of the Raman G-peak map in a). The blue (in c)) and green (in d)) rectangles mark the areas of the amplitude images shown in the main text in Figure 1g at an hBN edge and 1d of the uncovered TLG, also shown in e) and f), respectively. The scale bars in b)-d) correspond to 5 µm and in e) and f) to 1µm.*

We support the assignment of the stacking orders in the investigated TLG flake in Figure 1 in the main text with Raman spectroscopy[1,2]. A map of the integrated G-peak intensity from 0.196 to 0.197 eV is shown in Figure S1a (top). The darker areas (labeled '1' and '2') correspond to TLG covered with hBN, while the lighter parts are uncovered TLG. In both, covered and uncovered TLG, we observe distinct domains. Raman spectra around the 2D peak (shown in the bottom) allow to assign these different areas to ABA and ABC stacking orders as the 2D peak shows a sharp side peak and an enhanced shoulder for ABC stacked TLG. [1,2]

In the AFM topography image (Figure S1b) we observe a triangular-shaped area of high topography with a step-like height profile in the upper right. This triangular-shaped area corresponds to the hBN flake covering the TLG. The s-SNOM amplitude images of the same sample taken at 0.181 and 0.176 eV (Figure S1c and S1d) reveal a high s-SNOM amplitude in the area of the hBN flake due to hBN's high permittivity at the lower bound of the upper reststrahlenband. At the edges of the hBN flake, we observe bright fringes arising from the interference of polaritons in the heterostructure as discussed in Figure 2 in the main text.

At 0.181 eV (c.f. Figure S1c) the heterostructure shows a lower s-SNOM amplitude in the upper part of the hBN-covered TLG. Supported by the Raman G-peak map showing a lower G-peak energy in this area, we assign the upper part of the hBN-covered TLG to the hBN-covered ABC TLG. The rest of the hBN flake covers ABA TLG. The area of uncovered TLG shows areas of two distinct s-SNOM amplitudes. These areas agree with the areas that show distinct G-peak energies on the right of the Raman map in Figure S1a. Thus, we assign the uncovered areas with lower s-SNOM amplitude to ABA TLG and those with higher s-SNOM amplitude to ABC TLG.

At 0.176 eV (c.f. Figure S1d) we do not observe areas of different s-SNOM amplitudes in the area of the hBN-flake. In the uncovered TLG area we observe the same TLG domains as at 0.181 eV, but with the ABA TLG showing a higher s-SNOM amplitude and the ABC TLG showing a lower s-SNOM amplitude. The inversed contrast arises from the energy-dependent conductivities of the stacking orders that also lead to an energy-dependent s-SNOM amplitude contrast between the stacking orders.

The blue and green rectangles in Figure S1c and S1d, also shown in Figure S1e and S1f, respectively mark the areas of the s-SNOM amplitude images shown in Figure 1g and 1d in the main text, respectively.

## Supplementary Note 2: Stacking-dependent polaritons in hBN-covered TLG

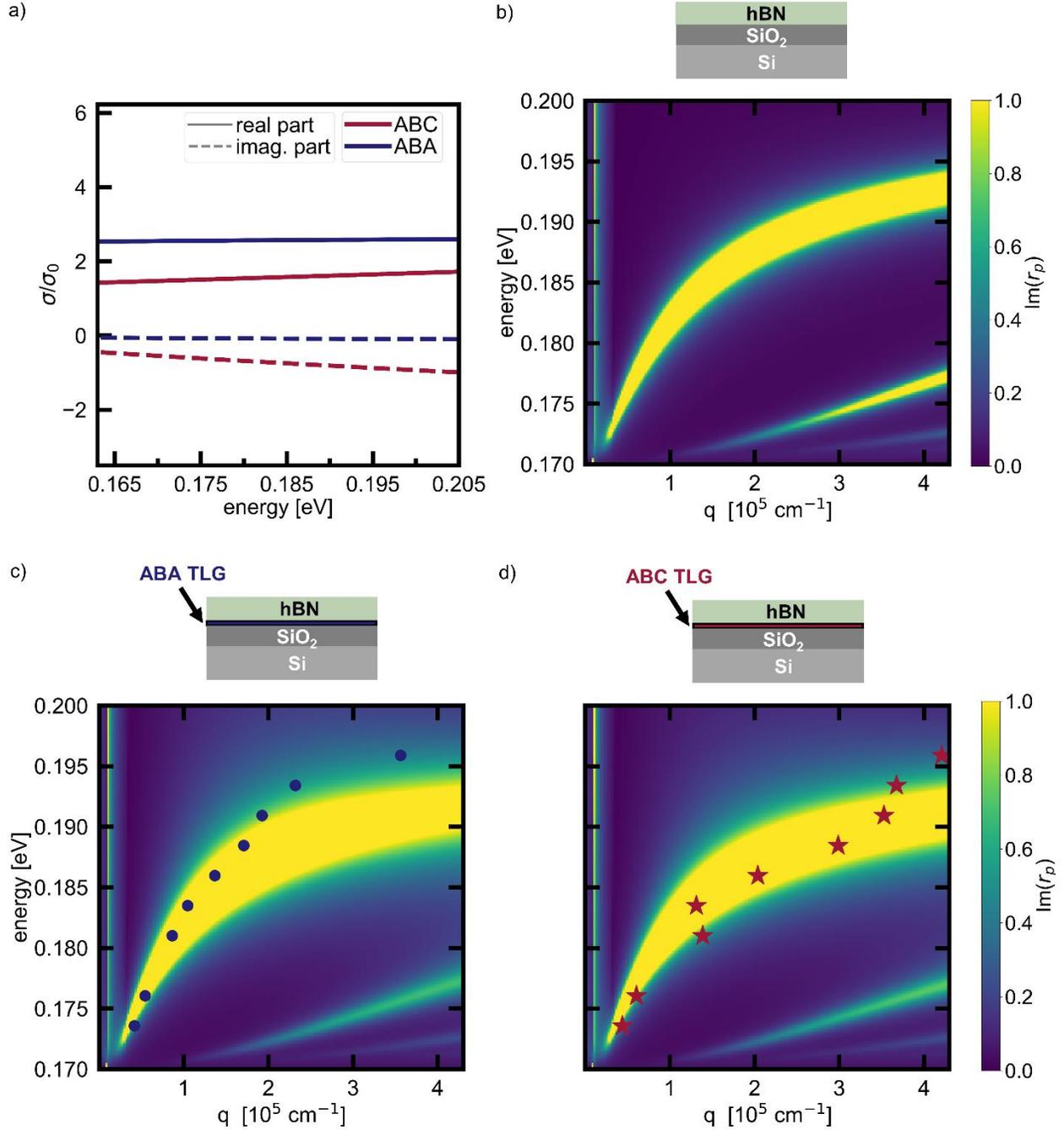

*Figure S2: Hybridization of the plasmon polaritons in TLG and the hyperbolic phonon polaritons in hBN leading to the stacking-dependent dispersion.* a) Real (solid line) and imaginary (dashed line) part of the optical conductivities of ABA (blue) and ABC (red) TLG in the investigated energy regime. b) Modeled dispersion of the hyperbolic phonon polaritons in a 33 nm thick hBN slab on the $SiO_2$-Si. c) and d) Comparison of the experimentally obtained (blue dots in c) and red stars in d)) and modeled dispersion of the hyperbolic phonon plasmon polaritons for the heterostructure with ABA TLG and ABC TLG, respectively.

Figure S2a shows a zoom-in of the optical conductivities of ABA and ABC TLG shown in Figure 1a and 1b in the investigated energy regime. In this energy regime ABC TLG's optical conductivity shows a higher magnitude in the imaginary part and a lower real part than ABA TLG's conductivity.

The modeled dispersion relation (color plot of the imaginary part of the reflection coefficient) is calculated for an hBN slab on the $SiO_2$-Si substrate (Figure S2b) and the two heterostructures (hBN ABA TLG in Figure S2c and hBN ABC TLG in Figure S2d) using the transfer matrix method.[3] The dispersion is calculated for the sample stacks sketched at the top of S2b-S2d, respectively. The $SiO_2$ is assumed to have a thickness of 90 nm. The hBN thickness is 33 nm. The permittivity of hBN is shown in Figure 1b and the optical conductivities of the TLG stacking orders are in Figures 1d and 1e.

For a pure hBN slab (Figure S2b) we observe a pronounced principal phonon polariton branch and two higher-order phonon polariton modes at high q-vectors and low energies. The higher-order modes arise from the waveguide-like propagation of the hBN phonon polaritons. In the heterostructures with hBN on top of ABA and ABC TLG, the modes are broadened and shifted to slightly lower energies. Comparing the two heterostructures we observe that the principal branch of the ABC heterostructure is shifted to slightly higher wavevectors compared to that of the ABA heterostructure. In the heterostructure of hBN on top of TLG, the phonon polaritons couple to the surface plasmon polaritons in TLG. Therefore, we observe hybridized modes in the dispersion of the heterostructure. Since the two stacking orders of TLG differ in their optical conductivities, the TLG surface plasmon polaritons are stacking-dependent[4] and we observe differently hybridized modes in the heterostructures with hBN on top of ABA and on top of ABC TLG.

For both heterostructures, the experimental data agree with the principal mode in the modeled dispersion. The experimental dispersion of the ABA heterostructure lies at slightly higher energies than the modeled dispersion for energies above 0.19 eV whereas the experimental dispersion of the ABC heterostructure agrees with the modeled dispersion up to an energy of 0.193 eV.